\documentstyle[12pt]{article}
\newlength{\extraspace}
\newlength{\extraspaces}
\newcommand{\be}{\begin{equation}
\addtolength{\abovedisplayskip}{\extraspaces}
\addtolength{\belowdisplayskip}{\extraspaces}
\addtolength{\abovedisplayshortskip}{\extraspace}
\addtolength{\belowdisplayshortskip}{\extraspace}}
\newcommand{\ee}{\end{equation}}

\newcommand{\ba}{\begin{eqnarray}
\addtolength{\abovedisplayskip}{\extraspaces}
\addtolength{\belowdisplayskip}{\extraspaces}
\addtolength{\abovedisplayshortskip}{\extraspace}
\addtolength{\belowdisplayshortskip}{\extraspace}}
\newcommand{\ea}{\end{eqnarray}}

\newcommand{\il}{\lambda_{{}_{{}_{\!\!\!\!\scriptstyle{i}}}}}

\newcommand{\bl}{\lambda_{{}_{{}_{\!\!\!\!\scriptstyle{0}}}}}

\newcommand{\al}{\lambda_{{}_{{}_{\!\!\!\!\scriptstyle{a}}}}}
\newcommand{\kl}{\lambda_{{}_{{}_{\!\!\!\!\scriptstyle{k}}}}}

\newcommand{\nonu}{\nonumber \\[.5mm]}
\newcommand{\A}{&\!\!\!}

\newcommand{\newsection}[1]{
\vspace{7mm}
\pagebreak[3]
\addtocounter{section}{1}
\setcounter{equation}{0}
\setcounter{subsection}{0}
\setcounter{footnote}{0}
\begin{center}
{\large {\bf \thesection. #1}}
\end{center}
\nopagebreak
\medskip
\nopagebreak
\hspace{3mm}}

\begin{document}


\begin{large}

\centerline{\bf COSMOLOGICAL APPLICATION OF}

\centerline{\bf THE NEW GENERAL RELATIVITY} 

\end{large}

\hspace{2cm}


\centerline{F.I. $Mikhail^{*}$, M.I. $Wanas^{**}$, and G.G.L. 
$Nashed^{*}$} 
\centerline {{*} Mathematics Department, Faculty of Science, Ain Shams 
University, Cairo, Egypt}

\centerline{ {**} Astronomy Department, Faculty of Science, Cairo University, 
Giza, Egypt}


\thispagestyle{empty}

\hspace{2cm}
\\
\\
\\
\\
\\

The field equations of the new general relativity constructed by Hayashi and 
Shirafuji (1979), have been applied to two different geometric structures, 
given by Robertson (1932), in the domain of cosmology. 
In the first application a family of
models, involving two of the parameters characterizing the field equations of
 the new general relativity, is obtained. In the second application the models
 obtained are found to involve one parameter only. The cosmological parameters
in both applications are calculated and some cosmological problems are discussed in comparison with the corresponding results of other field theories.

\newpage
\newsection{Introduction}

Standard Big Bang model suffers from some difficulties. Several trials 
have been made to overcome these difficulties [1--3]. These trials can be 
classified into two classes. The first class considers the modification of the
 model itself 
\cite{L9}. The second class considers the modification of the general 
relativity (GR) [4--5]. In this respect there is a general belief that
 the geometrical  structure upon which GR is based (Riemannian geometry), 
should
be replaced by a more wider structure. Some authors believe that the absolute
 parallelism (AP) space is a good candidate for this purpose [6--11]. Some
of these trials have been applied to cosmological problems [1--2], while
 others have not been applied, so far, in this domain.

It is the purpose of this paper to apply the new general relativity (NGR), 
constructed by Hayashi and Shirafuji (1979), in the domain of cosmology and
 to compare the obtained results with those of the previous theories. In 
section 2 we give a brief review of the geometric structure upon which the
NGR based on it. In section 3, we are going to give the two tetrad spaces
satisfying the cosmological principle, i.e, the homogeneity and isotropy of the
 universe, constructed by Robertson (1932), to 
apply them to the field equations of the NGR. In section 4 a brief review of 
the NGR is discussed. In section 5, application of the field equations of the 
NGR to the two tetrad spaces given in section 3. In section 6 
we compare the results with different field theories based on the AP-space.

\newsection{AP-Spaces}

The AP-space is a 4-dimensional vector space, in which every point is labelled
 by a set of 4-independent variables $x^{\nu}$ $(\nu=0,1,2,3)$. We associate
 with each point of this space 4-linearly independent contravariant vectors 
denoted by ${\il}^\mu$ where $(\mu=0,1,2,3)$ indicates the coordinate component
 and
 $(i=0,1,2,3)$ indicates the vector number. Each component of ${\il}^\mu$ is 
assumed to
 be a singal-valued function of $x^\nu$ and differentiable to any required 
order. The normalized cofactor of ${\il}^\mu$ in the determinate 
$\lambda$=$\| {\il}^\mu \| \neq 0$ 
is denoted by ${\il}_\mu$ and is defined by
${\il}_\mu=\displaystyle {1 \over \lambda} 
\left( cofactor \quad of \quad {\il}^\mu \quad on \quad \| {\il}^\mu \| 
\right)$.
It can be easily shown that ${\il}_\mu$ are covariant components of the tetrad 
and 
\be
{\il}^\mu {\il}_\nu={\delta^\mu}_\nu,
\ee
\be
{\il}_\mu {\kl}^\mu=\delta_{i k},
\ee
where ${\delta^\mu}_{\nu}$ are the ordinary Kronecker delta. Greek indices 
(world) are
 used to indicate the coordinate components and are written in a covariant
 or contravariant position, i.e, ${\il}^\mu$ or ${\il}_\mu$. 
Latin indices are used to indicate the
 number of the vector which are written in a lower position only and are called  "mesh indices". A summation convention will be applied to the mesh indices
as well as to the world one. Using the tetrad vector, we can define the 
following second order symmetric tensor:

\be
g_{\mu \nu}={\il}_\mu {\il}_\nu,
\ee
\be
g^{\mu \nu}={\il}^\mu {\il}^\nu,
\ee
thus, $g_{\mu \nu}g^{\nu \sigma}={\delta^\mu}_\sigma$.

It has been shown \cite{M2} that $(g)^{1/2}=\lambda$ where $g=\|g_{\mu \nu} \|$. 
We can always define
 a Riemannian space, associated with the AP-space, by taking the symmetric 
tensor (2.3) to play the role of the metric tensor, viz.
\be
ds^2=g_{\mu \nu} dx^\mu dx^\nu.
\ee
It is obvious from (2.5) that the associated Riemannian space will have a +ve 
definite metric. To compare the results of any theory, based on the AP-space, 
with the results of GR, the metric of the associated Riemannian space should 
have the Lorentz signature (indefinite). In this case the metric will take the 
form:
\be
ds^2={g^*}_{\mu \nu} dx^\mu dx^\nu,
\ee
where ${g^*}_{\mu \nu}=e_i {\il}_\mu {\il}_\nu$, $e_i$ is the Levi-Civita's indicator $(-1,+1,+1,+1)$.
 Some authors, instead of using Lorentz indicator $e_i$ in their applications,
 prefer to insert Lorentz signature by multiplying the zeroth vector by 
$(i=(-1)^{1/2})$ \cite{S8}. The symmetric tensor (2.3) and its conjugate (2.4) 
can be used to raise and lower world indices in the usual manner.

In the AP-space we can define two different connections:\\
(i) A symmetric connection $\{^{\alpha}_{\mu \nu} \}$ defined in terms of the symmetric tensor
$g_{\mu \nu}$ given by (2.3) in the standard way, i.e,
\be
\{^{\alpha}_{\mu \nu} \}={1 \over 2}g^{\sigma \alpha} \left( 
g_{\mu \sigma,\nu}+g_{\nu \sigma,\mu}-g_{\mu \nu,\sigma} \right).
\ee
(ii) A non symmetric connection ${\Gamma^\alpha}_{\mu \nu}$ defined in terms
 of the tetrad vectors directly as a consequence of the AP-condition \cite{M2}
\be
{\Gamma^\alpha}_{\mu \nu}={\il}^\alpha {\il}_{\mu,\nu}=-{\il}_\mu 
{{\il}^\alpha}_{,\nu}
\ee
Consequently, three different types of the absolute derivatives can be defined:
\be
{A^\mu}_{; \nu}={A^\mu}_{,\nu}+\{^{\mu}_{\alpha \nu} \} A^\alpha,
\ee
\be
{A^{\mu}_{+}}_{|\nu}={A^\mu}_{,\nu}+{\Gamma^\mu}_{\alpha \nu}A^\alpha,
\ee
\be
{A^{\mu}_{-}}_{|\nu}={A^\mu}_{,\nu}+{\Gamma^\mu}_{\nu \alpha}A^\alpha,
\ee

where
${A^\mu}_{,\nu}=\displaystyle{\partial A^\mu \over \partial x^\nu}$. 
We can define the following third order tensors:
\ba
{\Lambda^\alpha}_{\mu \nu}\A=\A{\Gamma^\alpha}_{\mu \nu}-{\Gamma^\alpha}_{\nu \mu}
=-{\Lambda^\alpha}_{\nu \mu},\nonu
{\gamma^\alpha}_{\mu \nu}\A=\A{\Gamma^\alpha}_{\mu \nu}- \{^{\alpha}_{\mu \nu}
 \}\nonu
{\Delta^\alpha}_{\mu \nu}\A=\A{\gamma^\alpha}_{\mu \nu}+
{\gamma^\alpha}_{\nu \mu}
\ea
It can be easily shown that \cite{M2,M7}
\be
{\gamma^\alpha}_{\mu \nu}={\il}^\alpha {\il}_{\mu;\nu}=
-{\il}_\mu {{\il}^\alpha}_{;\nu},
\ee
\be
\gamma_{\alpha \mu \nu}=-\gamma_{\mu \alpha \nu}.
\ee

It is clear that the skew part and the symmetric 
part of ${\Gamma^\alpha}_{\mu \nu}$ and ${\gamma^\alpha}_{\mu \nu}$ are 
respectively given by
\be
{\Gamma^\alpha}_{[\mu \nu]}={\gamma^\alpha}_{[\mu \nu]}={1 \over 2}
{\Lambda^\alpha}_{\mu \nu},
\ee
\be
{\gamma^\alpha}_{(\mu \nu)}={1 \over 2}{\Delta^\alpha}_{\mu \nu}.
\ee
The tensor ${\Lambda^\alpha}_{\mu \nu}$ is called the torsion tensor and 
${\gamma^\alpha}_{\mu \nu}$ is the contortion tensor. Both can be related by
\be
\gamma_{\alpha \mu \nu}=
{1 \over 2} \left( \Lambda_{\alpha \mu \nu}+\Lambda_{\nu \mu \alpha}
-\Lambda_{\mu \alpha \nu} \right).
\ee
Contracting any one of the three tensors (2.12) by setting $\alpha=\nu$ we get
 the basic vector
\be
C_\mu={\Lambda^\alpha}_{\mu \alpha}={\gamma^\alpha}_{\mu \alpha}
={\Delta^\alpha}_{\mu \alpha}.
\ee
This is the same vector denoted by Hayashi and Shirafuji \cite{HT7} by $v_\mu$ with
 a negative sign. They introduced another third order tensor given by:
\be
t_{\alpha \mu \nu}={1 \over 2}(\Lambda_{\alpha \mu \nu}+
\Lambda_{\mu \alpha \nu})-{1 \over 6}(g_{\nu \alpha}C_\mu+g_{\mu \nu}C_\alpha)
+{1 \over 3}g_{\alpha \mu}C_\nu,
\ee
and the axial vector
\be
a_\mu={1 \over 6} \epsilon_{\mu \nu \rho \sigma}\Lambda^{\nu \rho \sigma},
\ee
where
\be
\epsilon_{\mu \nu \rho \sigma}=\sqrt{-g}\delta_{\mu \nu \rho \sigma}
\ee
with $\delta_{\mu \nu \rho \sigma}$ completely skew symmetric and normalized
as $\delta_{0123}=-1$. The tensor defined by (2.19) has the following 
properties:
\ba
t_{\alpha \mu \nu}\A=\A t_{\mu \alpha \nu},\nonu
 g^{\mu \nu}t_{\alpha \mu \nu}\A=\A 0,\nonu
 t_{\alpha \mu \nu}+t_{\mu \nu \alpha}+t_{\nu \alpha \mu}\A= \A 0.
\ea

\newsection{Geometric Structures}

In order to examine the NGR, with respect to the cosmological problem, we are 
going to use the two tetrad spaces given by Robertson \cite{R3} and satisfying
the cosmological principle (homogeneity and isotropy of the universe). The two structures have the same associated Riemannian space characterized by the well
 known Robertson-Walker metric. These two geometric structures are defined by 
the following tetrads (written in the spherical polar coordinate, i.e., 
 $x^0=t$, $x^1=r$, $x^2=\theta$, $x^3=\phi$) viz
\be
 {\il}^\mu  = 
\left(
\matrix{
\sqrt{-1} & 0 & 0 & 0 \vspace{3mm} \cr 
0 &\left( \displaystyle{L^+ sin\theta cos\phi \over 4R} \right) & 
\left( \displaystyle{L^- cos\theta cos\phi-4k^{1 \over 2}r sin\phi \over 4rR} 
\right) & 
-\left( \displaystyle{L^- sin\phi+4k^{1 \over 2} r cos\theta cos\phi \over 
4rRsin\theta} \right) \vspace{3mm} \cr 0 & 
\left( \displaystyle{L^+ sin\theta sin\phi \over 4R} \right) &
\left( \displaystyle{L^- cos\theta sin\phi-4k^{1 \over 2}r cos\phi \over 
4rR} \right) & 
\left( \displaystyle{L^- cos\phi-4k^{1 \over 2} r cos\theta sin\phi \over
 4rRsin\theta} \right) \vspace{3mm} \cr 0 & 
\displaystyle{L^+ cos\theta \over 4R} & 
-\displaystyle{L^- sin\theta \over 4rR} & \displaystyle{k^{1 \over 2} \over 4R}
 \cr}\right),
\ee
and
\be
 {\il}^\mu  = 
\left(
\matrix{
\sqrt{-1}\displaystyle{L^-  \over L^+} & -\displaystyle{k^{1 \over 2}r \over R} & 0 & 0 \vspace{3mm} \cr 
\sqrt{-1} \left( \displaystyle{4k^{1 \over 2}r sin\theta cos\phi \over L^+} 
\right)
 & \left( \displaystyle{L^- sin\theta cos\phi \over 4R} \right) & 
\left( \displaystyle{L^+ cos\theta cos\phi \over 4rR} \right) & 
-\left( \displaystyle{L^+ sin\phi \over 4rRsin\theta} \right) \vspace{3mm} \cr
\sqrt{-1} \left( \displaystyle{4k^{1 \over 2}r sin\theta sin\phi \over L^+} 
\right) & 
\left( \displaystyle{L^- sin\theta sin\phi \over 4R} \right) &
\left( \displaystyle{L^+ cos\theta sin\phi \over 4rR} \right) & 
\left( \displaystyle{L^+ cos\phi \over 4rRsin\theta} \right) \vspace{3mm} \cr 
\sqrt{-1} \left( \displaystyle{4k^{1 \over 2}r cos\theta  \over L^+} \right)  & \left( \displaystyle{L^- cos\theta \over 4R} \right) & 
-\left( \displaystyle{L^+ sin\theta \over 4rR} \right) & 0
 \cr}\right),
\ee
where $L^{\pm }=4 \pm kr^2$,  k is the curvature constant $(k=-1,0,+1)$, and 
R(t) is an unknown function of t. The Riemannian space 
associated with the tetrads (3.1), (3.2) will have the same Robertson-Walker
 metric:
\be
ds^2=-dt^2+{16 R^2(t) \over L^{+2}} \left( dr^2+r^2(d\theta^2+sin^2\theta 
d\phi^2) \right)
\ee

\newpage
\newsection{Hayashi-Shirafuji Theory}

In 1979 Hayashi and Shirafuji constructed a theory which they called "New
 General Relativity", NGR. They have used an AP-space for its formulation,
 and a variational principle to derive its field equations. They were able to
derive a set of field equations involving three different parameters. They took
 the Lagrangain density in the form
\be
{\cal L}= {\lambda \over 2\kappa} 
\left[R+2{\overline d_1}(t^{\mu \nu \lambda}t_{\mu \nu \lambda})+
2{\overline d_2}(v^{\mu} v_{\mu})+2{\overline d_3}(a^{\mu}a_{\mu}) \right],
\ee
where ${\overline d_1}$, ${\overline d_2}$, and ${\overline d_3}$ are 
dimensionless parameters of the theory. They have chosen their Lagrangian such
that it will be invariant under the following:\\
a) the group of general coordinate transformations,\\
b) the group of global Lorentz transformations,\\
c) the parity operation, i.e.,
${\bl}^\mu \rightarrow {\bl}^\mu$, and ${\al}^\mu \rightarrow -{\al}^\mu$,\\
where $a=1,2,3$, R is the Ricci scalar tensor and $t_{\alpha \mu \nu}$, $C_\mu$
, $a_\mu$, are the tensors defined in section 2. By applying a variational 
principle to the above Lagrangian they \cite{HT7} were able to obtained the 
field equations in the form:

\be
G^{\mu \nu}+2{\kappa}{D}_\lambda F^{\mu \nu \lambda}-
2{\kappa} C_\lambda F^{\mu \nu \lambda}+2{\kappa}H^{\mu \nu}
-{\kappa} g^{\mu \nu}L'={\kappa}T^{\mu \nu},
\ee
where $G^{\mu \nu}$ is the Einstein tensor defined by
\be
G^{\mu \nu}=R^{\mu \nu}-{1 \over 2}g^{\mu \nu} R
\ee
with $R^{\mu \nu}$ being the Ricci tensor, and
\ba
F^{\mu \nu \lambda}
 \A = \A {1 \over \kappa} \left[ {\overline d_1} \left(t^{\mu \nu \lambda}
-t^{\mu \lambda \nu} \right)-{\overline d_2} \left(g^{\mu \nu} C^\lambda
-g^{\mu \lambda} C^\nu \right)
-{{\overline d_3} \over 3} \epsilon^{\mu  \nu \lambda \rho} a_\rho \right]
=-F^{\mu \lambda \nu},\\
H^{\mu \nu} \A = \A \Lambda^{\rho \sigma \mu} 
 {F_{\rho \sigma}}^\nu - {1 \over 2} \Lambda^{\nu \rho \sigma} 
{F^\mu}_{\rho \sigma}=H^{\nu \mu},\\
{L'} \A = \A {{\cal L} \over \lambda},\\
T^{\mu \nu} \A = \A {1 \over \lambda} {\delta {\cal L}_M \over 
\delta {b^k}_\nu} b^{k \mu}.
\ea
Here ${\cal L}_M$ denotes the Lagrangian density of material fields and
$T^{\mu \nu}$ is the material energy-momentum tensor which is 
nonsymmetric in general.

It is well known that the conservation law in GR is given by
\be
{{T_{GR}}^{(\mu \nu)}}_{;\nu}= 0,
\ee
where ${T_{GR}}^{(\mu \nu)}$ is the symmetric material energy-momentum  tensor
of GR and semicolon denotes the covariant derivative with respect to the 
Christoffel symbol. This law does not follow from (4.2), however.
Instead, they \cite{HT7} derive the response equation 
\be
{T^{\mu \nu}}_{;\nu}=\gamma^{\nu \lambda \mu}T_{[\nu \lambda]}.
\ee
The antisymmetric part $T^{[\mu \nu]}$ is due to the contribution from the
intrinsic spin of fundamental spin-$1/2$ particles. For macroscopic test
particles  employed in terrestrial experiments, effects due to intrinsic spin 
can be ignored, and hence their energy-momentum tensor can be supposed to be
symmetric and satisfy (4.8). Accordingly, the equation of motion for
macroscopic test particles is the geodesic equation of the metric.

\newsection{Cosmological Models}

In the present section we are going to apply (4.2) together with the 
response equation (4.9) to the two geometric structures given in section 3. In 
order to compare the models that will arise from the NGR with observation we
 are going to use the following parameters \cite{N8}.\\
The Hubble's parameter, $H_0$=$\displaystyle{\dot{R}(t_0) \over R(t_0)}$=
$\displaystyle{{\dot{R}}_0 \over R_0}$= $100h_0 kms^{-1} Mpc^{-1}$, where 
0.5 $\leq h \leq1$,\\
the deceleration parameter, $q_0$=-$\displaystyle{\ddot{R_0} \over R_0}
{H_0}^{-2},$\\
the denisty parameter, $\sigma_0$=$\displaystyle{\rho_0 \over \rho_c}$,\\
where $\rho_c$ is the critical denisty defined by, $\rho_c$=
${3H^2/\kappa}$ \cite{N8}, $\rho_0$ is the proper denisty
 given by the model, $\kappa$ is the Einstein constant (=8$\pi$ in 
relativistic units), and the dot denotes differentiation with respect to the 
time t. For the present application we are going to use the material energy
 tensor of the perfect fluid with the following non-vanishing components
\be
{T^0}_0=\rho_0, \quad {T^1}_1={T^2}_2={T^3}_3=-p_0.
\ee
In the dust case the equation of the state will be
\be
p_0=0.
\ee
In the case of radiation, the equation of state will be
\be
p_0={1 \over 3}\rho_0.
\ee

\subsection{\it {MODELS RESULTING FROM THE FIRST STRUCTURE.}}
Applying the field equations (4.2) to the tetrad (3.1), we have found that the
 skew part of the L.H.S. of (4.2) is identically zero, hence $T^{[\mu \nu]}=0$.
 In this case the field equations (4.2) will give rise to the two differential
equations:
\ba
Y \displaystyle{\dot{R^2} \over R^2}+B \displaystyle{k \over R^2}=
{\kappa \over 3}\rho_0,\nonu
2Y \displaystyle{\ddot{R} \over R}+Y \displaystyle{\dot{R^2} \over R^2}+
B \displaystyle{k \over R^2}=-\kappa p_0,
\ea
where the new parameters Y and B are given by $Y=(1-3{\overline d_2})$,
$B=(1+4/3{\overline d_3})$. Also since $T^{[\mu \nu]}=0$, the general law of 
the response equation of NGR will reduce to the standard conservation law of GR
giving rise in the present application, to
\be
R \displaystyle{d\rho_0 \over dR}=-3(\rho_0+p_0).
\ee
The solutions of (5.4) and (5.5) in the two cases of the physical interest 
given by (5.2) and (5.3) are summarized in the following table

\bigskip

\centerline{Table(I) Solution of the field equations}
\begin{tabular}{|c|c|c|}\hline 
 {Curvature Constant} & {Dust Case} & {Radiation Case} \\ \hline
{$k=+1$} & {$R=\beta(1-cosw)$ \quad $t=\left( {Y \over B} \right)^{1 \over 2}
\beta(w-sinw)$} & 
{$R(t)$}={$ \displaystyle{ \left( \gamma^2-[t \delta^2-\gamma^2]^2 \right)
^{1 \over 2}} \over \delta$} \\ \hline
{$k=0$} &  {$R(t)$}=
{$\left( \displaystyle{9B \beta \over 2Y} \right)^{1 \over3}$}
{$t^{2 \over 3}$} & {$R(t)$}={$(2\gamma)^{1 \over 2}$ $t^{1 \over 2}$} 
\\ \hline
{$k=-1$} & {$R$}={$\beta(coshw-1)$ $t=
\left( \displaystyle{Y \over B} \right)^{1 \over 2} \beta(sinhw -w)$} & 
{$R(t)$}={$\displaystyle{ \left( [t \delta^2+\gamma^2]^2 -\gamma^2 \right)^
{1 \over 2}} \over \delta$} \\ \hline
\end{tabular}
where $\gamma^2$=${\kappa M/3Y}$, 
$\delta^2$=${B/Y}$, 
$\beta$=${4\pi N/3B}$, M and N being  constants of 
integration resulting from (5.5). These solutions involve the two parameters 
${\overline d_2}$,
${\overline d_3}$ characterizing the field equations of the NGR.

Using the definitions of the parameters listed above we get the following values for the solutions given in table I \\
\newpage

\begin{center}
 \centerline{Table(II) Summary of The Cosmological Parameters}
\begin{tabular}{|c|l|c|c|}\hline 
\multicolumn{2}{|c|} {Parameters} & {Dust Case} & {Radiation Case} \\ \hline
 & $k=+1$ & 
 $\displaystyle{1 \over 1+\cos w}$ & 
$\displaystyle{\gamma^2 \over (t \delta^2 -\gamma^2)^2}$ \\
$q_0$ & $k=0$ & $\displaystyle{1 \over 2}$ & $1$ \\
 & $k=-1$ &  $\displaystyle{1 \over 1+cosh w}$ & 
$\displaystyle{\gamma^2 \over (t \delta^2 +\gamma^2)^2}$ \\ \hline
 & $k=+1$ & 
 $\left( {B \over Y} \right)^{1/2}  \displaystyle{sin w \over R_0 (1-\cos w)^2}$ & $-\displaystyle{(\gamma-t) \over (2 t \gamma -t^2)}$ \\
$H_0$ & $k=0$ & 
 ${2 \over 3} {t_0}^{-1}$ & ${1 \over 2}{t_0}^{-1}$ \\
 & $k=-1$ & 
 $\left( {B \over Y} \right)^{1/2} \displaystyle{sinh w \over R_0 (cosh w-1)^2}$ & $\displaystyle{(\gamma+t) \over (2 t \gamma +t^2)}$  \\ \hline
& $k=+1$ & 
 $\displaystyle{2 Y \over 1+\cos w}$ & 
$\displaystyle{\gamma^2 Y \over 2(t \delta^2 -\gamma^2)^2}$ \\
$\sigma_0$ & $k=0$ & Y & $\displaystyle{Y \over 2}$ \\
 & $k=-1$ & 
 $\displaystyle{2 Y \over 1+cosh w}$ & 
$\displaystyle{\gamma^2 Y \over 2(t \delta^2 +\gamma^2)^2}$ \\ \hline
\end{tabular}
\end{center}
Now we are going to discuss some defects of the above models

{\underline {\bf particles horizon.}}\\
We are going to examine whether or not the resulting world models involve
 particle horizones. To do so we have to recall the metric of the Riemannian
 space associated with these models as given by (3.3), i.e.,
\be
ds^2=-dt^2+{16R^2(t) \over L^{+2}} [dr^2+r^2(d\theta^2+sin^2\theta d\phi^2)],
\ee
according to this metric, the coordinate radial distance of any particle of the
 model from the origin of coordinates is given by
\be
r=\int{dr \over 4+kr^2},
\ee
hence the proper distance of this particle from the origine is given by
\be
S=4R_0 r=4R_0\int{dr \over 4+kr^2}.
\ee
To find out the limiting proper distance up to which observations can be 
carried out, we note that for a radial ray of light, we get
\be
\int_{t_{1}}^{t{0}}{dt \over 4R}=\int_0^{r}{dr \over 4+kr^2}.
\ee
${\underline {\bf For k=+1:}}$ \\
\be
r=2tan \left[ \left( {Y \over B} \right)^{1 \over 2} {(w_0-w_1) \over 2} 
\right],
\ee
where $w_0=cos^{-1} \left \{ {(1-q_0)/q_0} \right \},$
$w_1=cos^{-1} \left \{1- {(2q_0-1)/q_0(1+Z)} \right \}$.
We can see from (5.10), that r has a finite value if $Z\rightarrow \infty$. 
For simplicity we call this finite value as $r_f$ and integrate (5.8) from 0 to $r_f$, we finally get
\be
S=2 R_0 tan^{-1}{r_f \over 2} \qquad for \quad k=+1.
\ee
Similarly we can get
\ba
S \A = \A R_0 r_f \qquad for \quad k=0, \nonu
S \A =\A 2R_0 tanh^{-1}{r_f \over 2} \qquad for \quad k=-1,
\ea
from (5.4) and the definition of the parameters we can get the relation
\be
2q_0-1={k \over {R_0}^2{H_0}^2} {B \over Y}={k q_0 R_0 \over \beta},
\ee
which will give rise to
\be
R_0=({B \over Y})^{1 \over 2} {1 \over H_0} \sqrt{{k \over (2q_0-1)}} \qquad
 for \quad k=\pm1,
\ee
and for $k=0$, $R_0$ is given by
\be
R_0= \left( {9B \beta \over 2Y} \right)^{1 \over 3} {t_0}^{2 \over 3}.
\ee
We see from (5.11), and (5.12) that S has a finite value as $Z\rightarrow 
\infty$ in the three cases. This means that all the above models contain particle
 horizons. However, it may be of interest to point out that as $Y\rightarrow 0$, i.e., $\kappa{\overline d_2} \rightarrow {1 \over 3}$, the horizon will be
 extended further and further and $S\rightarrow \infty$. This value is
calculated  in the case of dust. It can be shown that the same calculations 
could be drawn in the case of radiation.\\
{\underline {\bf Flatness problem.}}\\
>From (5.4) we have
\be
{\dot{R^2} \over R^2}={\kappa \over 3Y} \rho_0-\delta^2{k \over R^2},
\ee
which will give rise to
\be
{\sigma_0-1 \over \sigma_0}={3B \over \kappa}{k \over \rho_0 R^2},
\ee
it was shown \cite{T7}
\ba
{\sigma_0-1 \over \sigma_0}=\left\{
\matrix{ 
& \!\!\! 10^{-15}B \qquad {\rm at\ 1\ sec.} \hfill\cr
& \!\!\! 10^{-49}B \qquad {\rm at\ 10^{-35}\ sec.} \hfill\cr
& \!\!\! 10^{-57}B \qquad {\rm at\ 10^{-43}\ sec.} \hfill\cr
}\right.
\ea
It is clear from (5.18) that $\sigma_0 \sim 1$ when $k \sim 0$ and this means 
that the above models suffer from the flatness problem.

\subsection{\it {MODLES RESULTING FROM THE SECOND STRUCTURE.}}
Applying the field equations (4.2) to the tetrad (3.2), and using the same
procedure applied in the first application we get

\ba
\left( \displaystyle{\dot{R^2} \over R^2}+ \displaystyle{k \over R^2} \right)Y=
{\kappa \over 3}\rho_0,\nonu
\left( \displaystyle{2 \ddot{R} \over R}+ \displaystyle{\dot{R^2} \over R^2}+
 \displaystyle{k \over R^2} \right)Y=-\kappa p_0,
\ea
the solution of equation (5.19), in the cases mentioned by (5.2), (5.3) can be
 summarized
 in the following table

\centerline{Table(III) Solution of the field equations}
\begin{tabular}{|c|c|c|}\hline 
 {Curvature Constant} & {Dust Case} & {Radiation Case} \\ \hline
{$k=+1$} & {$R=\alpha(1-cosw)$ \quad $t=\alpha(w-sinw)$} & 
{$R(t)$}={${(2 \gamma t-t^2)^{1 \over 2}}$} \\ \hline
{$k=0$} &  {$R(t)$}= {$\left( \displaystyle{9 \alpha \over 2Y} \right)^{1 \over3}$}
{$t^{2 \over 3}$} & {$R(t)$}={$(2\gamma)^{1 \over 2} t^{1 \over 2}$} 
\\ \hline
{$k=-1$} & {$R$}={$\alpha(coshw-1) \quad t=\alpha(sinhw -w)$} & 
{$R(t)$}={${(2 \gamma t+t^2)^{1 \over 2}}$} \\ \hline
\end{tabular}

where $\alpha$=${4\pi/3Y}$. These solutions involve only
one parameter ${\overline {d_2}}$, characterizing the field equations of NGR.

Using the definitions of the parameters listed above we get the following values for the solutions given in table IV
\newpage
 \centerline{Table(IV) Summary of The Cosmological Parameters}
\begin{center}
\begin{tabular}{|c|l|c|c|}\hline 
\multicolumn{2}{|c|} {Parameters} & {Dust Case} & {Radiation Case} \\ \hline
 & $k=+1$ & 
 $\displaystyle{1 \over 1+\cos w}$ & 
$\displaystyle{\gamma^2 \over (t -\gamma)^2}$ \\
$q_0$ & $k=0$ & 
 $\displaystyle{1 \over 2}$ & $1$ \\
 & $k=-1$ & 
 $\displaystyle{1 \over 1+cosh w}$ & 
$\displaystyle{\gamma^2 \over (t +\gamma)^2}$ \\ \hline
 & $k=+1$ & 
 $\left( {B \over Y} \right)^{1/2} \displaystyle{sin w \over R_0 (1-cos w)^2}$ &  $-\displaystyle{(\gamma-t) \over (2 t \gamma -t^2)}$ \\
$H_0$ & $k=0$ & 
 $\displaystyle{2 \over 3} {t_0}^{-1}$ & 
$\displaystyle{1 \over 2}{t_0}^{-1}$ \\
 & $k=-1$ & 
 $\left( {B \over Y} \right)^{1/2}  
\displaystyle{sinh w \over R_0 (cosh w-1)^2}$ & 
$\displaystyle{(\gamma+t) \over (2 t \gamma +t^2)}$  \\ \hline
& $k=+1$ & 
 $\displaystyle{2 Y \over 1+\cos w}$ & 
$\displaystyle{\gamma^2 Y \over 2(t -\gamma)^2}$ \\
$\sigma_0$ & $k=0$ & Y & $\displaystyle{Y \over 2}$ \\
 & $k=-1$ & 
 $\displaystyle{2 Y \over 1+cosh w}$ & 
$\displaystyle{\gamma^2 Y \over 2(t +\gamma)^2}$ \\ 
\hline
\end{tabular}
\end{center}

We would like to point out that the models obtained here different from those
of GR. However, the models obtained from the second structure will reduce to
those of GR if $Y \rightarrow 0$.

{\underline {\bf particles horizon.}}\\
By the same procedure applied in the subsection 5.1 we can evaluate the proper
 distance for the present structure (in the case of dust)

\ba
S \A= \A R_0 tan^{-1}{r_f \over 2} \qquad for \quad k=+1,\\ \nonu 
S \A =\A R_0 r_f \qquad for \quad k=0, \\ \nonu 
S \A =\A R_0 tanh^{-1}{r_f \over 2} \qquad for \quad k=-1,
\ea
from (5.19) and the definition of the parameters we can get the relation
\be
2q_0-1={k \over {R_0}^2{H_0}^2}={k q_0 R_0 \over \alpha},
\ee
which will give rise to
\be
R_0= {1 \over H_0} \sqrt{{k \over (2q_0-1)}} \qquad for \quad k=\pm1,
\ee
and for $k=0$, $R_0$ is given by
\be
R_0=\left( \displaystyle{9\alpha  \over 2Y} \right)^{1 \over 3} {t_0}^
{2 \over 3}.
\ee
We see from (5.20) that S has a finite value as $Z\rightarrow 
\infty$ in the three cases. This means that all the above models contain 
particle horizons.

Also all the above models of the second application suffer from the flatness 
problem.
{\underline {\bf Singularity problem.}}\\
It clear from table (II) and (IV) that $H_0$ tends to infinity as 
$t\rightarrow 0$ in all the cases and this means that the denisty will be 
infinite. This means that the two applications suffer from the singularity
problem.

\newsection{\it {\bf Comparison With Other Field Theories.}}

Now we are going to compare the resultes of the present work with the results of other field theories, constructed using AP-space, to show the difference
between them in the cosmological application.

\centerline{Table(V) Comparison With Other Theories}

\begin{tabular}{|c|c|c|c|c|}\hline 
{Criterion}     & {GR (1916)} & {GFT(1977)} & {MTT (1978)} & 
{NGR (1979)} \\ \hline
Space-time      & Riemannian    & AP-Space         & AP-Space & AP-Space \\ 
                & Space         &                  &          &          
\\ \hline
Free Parameters &      No         &      No         &    One  & Three  \\ \hline%
Energy-Momentum & Symmetric Tensor & Symmetric & Symmetric  & Non-Symmetric \\               &                  & Geometric & Phenomol- & Phenomological \\
Tensor $T^{\mu \nu}$ &   & Tensor & ogical Tensor &  Tensor \\ \hline
Basic & Homogeneity & Homogeneity & Homogeneity & Homogeneity \\
Assumption & and Isotropy & and Isotropy & and Isotropy & and Isotropy \\ \hline%
Need for an        & Yes & No  & Yes & Yes \\
Equation of State  &     &     &      &    \\ \hline
Curvature Constant & $+1,0,-1$ & $-1$ & $+1,0,-1$ & $+1,0,-1$ \\
for Non-Static     &           &      &           &           \\      
Non-Empty Model    &           &      &           &           \\ \hline
Number of        & Many & One  & Many & Many \\
Allowed Models   &      &      &      &       \\ \hline
Theory & Macroscopic & Macroscopic & Macroscopic & Macroscopic and \\
       &             &             &             & Microscopic 
\\ \hline
Indicator for & Singularity & $\Lambda$ & Singularity& Singularity \\
Strong Field  &             &          &            &              \\ \hline
{Particle Horizon} & {Yes} & 
{No} & {Yes} & {Conditional} \\ \hline
{Flatness Problem} & {Yes} & 
{No} & {Yes} & {Yes} \\ \hline
{Singularity Problem} & {Yes} & 
{Yes} & {Yes} & {Yes} \\ \hline
\end{tabular}

Where $\Lambda={1 \over 2}(\sigma-{\overline {\omega}})$, 
${\overline {\omega}}={\gamma^\sigma}_{\mu \epsilon}
{\gamma^\epsilon}_{\sigma \nu}-{\gamma^\sigma}_{\nu \epsilon}
{\gamma^\epsilon}_{\sigma \mu}$, 
$\sigma_{\mu \nu}={\gamma^\sigma}_{\epsilon \mu}{\gamma^\epsilon}_{\sigma \nu}$



\newpage
\begin{thebibliography}{100}

\bibitem{S8} 
S$\acute{a}$ez, D., and De Juan, T.
(1984), {\it GRG}, {\underline {16}}, 5.\

\bibitem{W9}
Wanas, M.I.
(1989), {\it Astrophys. Space Sci.\ }, {\underline {154}} 165.\

\bibitem{L9}
Lind, A.D.
(1990), {\it "{\underline {Inflation and Quantum Cosmology}}"} Acadmic Press, 
Inc.\

\bibitem{H8}
Hoyle, F.
(1948), {\it M.\ N.\ R.\ A.\ S.\ }, {\underline {108}}, 372.\

\bibitem{H9}
Hoyle, F.
(1949), {\it M.\ N.\ R.\ A.\ S.\ }, {\underline {109}}, 365.\

\bibitem{E} 
Einstein, A.
(1929), {\it Sitz.\ der Preuss.\ Aksd.\ der wiss.\ }, {\underline 1}, 2.\

\bibitem{L5}
Levi-Civita, T.
(1950), {\it "A simplified presentation of Einstien's unified field equations"} London: Blackie and Son.\

\bibitem{M4}
Mikhail, F.I.
(1964), {\it Al Nuovo Cimento Series X }, {\underline {32}}, P. 886.\

\bibitem{MW7}
Mikhail, F.I., and Wanas, M.I.
(1977), {\it Proc.\ Roy.\ Soc.\ }, Lond.\ A {\underline {356}},  471.\

\bibitem{M7} 
M$\phi$ller, C.
(1978), {\rm Mat.\ Fys.\ Medd.\ Dan.\ Vid.\ Selsk.\ }, {\underline {39}}, 13.\

\bibitem{HT7} 
Hayashi, K., and  Shirafuji, T.
(1979) {\it Phys.\ Rev.\ D.\ }, {\underline {19}}, 3524.\

\bibitem{M2}
Mikhail, F.I.
(1962), {\it Ain Shams Sci.\ Bull.\ }, {\underline 6}, 87.

\bibitem{R3}
Robertson, H.P.
(1932), {\it Ann.\ Math.\ Princeton}, {\underline {33}}, 496.\

\bibitem{N8}
Narlikar, J.V.
(1983), {\it "{\underline {Introduction to Cosmology}}"}, Janes and Barlett.
Boston.\

\bibitem{T7}
Turner, M.S.
(1987), {\it GRG}, 11.\

\end{thebibliography}
\end{document}